\documentclass[
superscriptaddress,
final,
pre,
twocolumn,
showkeys,
floatfix,
]{revtex4-2}

\usepackage[utf8]{inputenc}
\usepackage{amsmath,amssymb,amsfonts,latexsym}
\usepackage{hyperref}
\usepackage{color}
\usepackage{graphicx}

\hypersetup{
    colorlinks=true,
    urlcolor=blue,
    linkcolor=blue,
}

\begin{document}
\title{Epistemological Fault Lines Between Human and Artificial Intelligence}

\author{Walter Quattrociocchi} 
\email{walterquattrociocchi@gmail.com}
\affiliation{Department of Computer Science,
Sapienza University of Rome, Rome, Italy}

\author{Valerio Capraro} 
\email{caprarovalerio@gmail.com}
\affiliation{Department of Psychology, University of Milan Bicocca, Milan, Italy}

\author{Matja{\v z} Perc} 
\email{matjaz.perc@gmail.com}
\affiliation{Faculty of Natural Sciences and Mathematics, University of Maribor, Maribor, Slovenia}
\affiliation{Community Healthcare Center Dr. Adolf Drolc Maribor, Maribor, Slovenia}
\affiliation{University College, Korea University, Seoul, Republic of Korea}
\affiliation{Department of Physics, Kyung Hee University, Seoul, Republic of Korea}

\date{\today}

\begin{abstract}
Large language models (LLMs) are widely described as artificial intelligence, yet their epistemic profile diverges sharply from human cognition. Here we show that the apparent alignment between human and machine outputs conceals a deeper structural mismatch in how judgments are produced. Tracing the historical shift from symbolic AI and information filtering systems to large-scale generative transformers, we argue that LLMs are not epistemic agents but stochastic pattern-completion systems, formally describable as walks on high-dimensional graphs of linguistic transitions rather than as systems that form beliefs or models of the world. By systematically mapping human and artificial epistemic pipelines, we identify seven epistemic fault lines, divergences in grounding, parsing, experience, motivation, causal reasoning, metacognition, and value. We call the resulting condition Epistemia: a structural situation in which linguistic plausibility substitutes for epistemic evaluation, producing the feeling of knowing without the labor of judgment. We conclude by outlining consequences for evaluation, governance, and epistemic literacy in societies increasingly organized around generative AI.
\end{abstract}
\maketitle

\section{Introduction}
\label{sec:intro}

The aspiration to build machines capable of mimicking or reproducing human thought long predates the advent of digital computers. Well before modern technology, myths and legends already testified to a fascination with artificial minds. In Greek mythology, for instance, Hephaestus is said to have crafted golden automatons that could move, speak, and reason, while in Jewish folklore the Golem appears as a man-made being animated through sacred letters and mystical rituals. By the late Middle Ages and the Renaissance, scholars such as Ramon Llull, and later Gottfried Wilhelm Leibniz, began to entertain the possibility of logical engines, devices that would manipulate symbols to carry out reasoning procedures. Leibniz’s proposal of a calculus ratiocinator, a universal symbolic language that could in principle resolve disputes through computation, anticipated both formal logic and, in the long run, theoretical computer science~\cite{lenzen2018leibniz}. In parallel, Enlightenment thinkers from Descartes to La Mettrie advanced mechanistic accounts of the mind, portraying human cognition as a system of interacting parts that might someday be reproduced artificially~\cite{gunderson1964descartes}.

The modern notion of machines that `think’ emerged in the mid-20th century, when developments in mathematical logic, computation, and electronics converged. In his seminal 1950 essay, Alan Turing proposed what he called the imitation game---now widely known as the Turing Test---as an operational criterion for intelligence~\cite{turing2007computing}. Crucially, Turing shifted the focus from defining thinking in the abstract to asking whether a machine’s outward behavior could be indistinguishable from that of a human interlocutor. Over the subsequent decade, artificial intelligence crystallized as a distinct research field, producing early systems for theorem proving, game playing, and symbolic problem solving~\cite{russell1995}. The long-standing philosophical speculation that machines might match human judgment---encompassing perception, reasoning, and even moral evaluation---thereby began to transform into the empirical and technological project whose consequences we are observing today.

In recent years, large language models (LLMs) have arguably been the most disruptive step in this trajectory~\cite{zhao2023survey}. State-of-the-art systems such as ChatGPT, Deepseek, Gemini, Llama, and Mistral now routinely clear the bar of the Turing Test, in some cases more reliably so than humans~\cite{jones2025large}. Unsurprisingly, a growing body of work has proposed LLMs as stand-ins for human participants in social science experiments~\cite{grossmann2023ai}, market and consumer research~\cite{brand2024using}, and a variety of applications in the healthcare, education, work and information domains~\cite{cascella2023evaluating,capraro2024impact,uygunilikhan_bktd25}, among others \cite{dwivedi2023opinion,ooi2025potential,budhwar2023human,raiaan2024review}. This comes on top of their widespread deployment for everyday tasks such as drafting and editing text, translation, summarization, and educational support~\cite{perc_it25}. Proponents have gone further, arguing that in many contexts LLMs may offer advantages over human samples, citing lower cost, greater scalability, and the ability to generate large volumes of synthetic data in domains where real-world data are limited or difficult to obtain~\cite{horton2023large, arora2025ai}.

However, serious concerns have already been voiced~\cite{dennett2023problem, gao2025take}, and a number of foundational issues remain unresolved. A central open question is how judgment itself is instantiated and operationalized in LLMs. These systems are rapidly becoming embedded in the processes by which societies filter, rank, and interpret information: assessing the credibility of news, proposing explanations, and assisting in decisions that hinge on evaluative judgments~\cite{deverna2024fact, kunz2024properties, ikeda2024inconsistent}. Yet the internal procedures by which they arrive at such judgments---and the extent to which these procedures align with, diverge from, or systematically distort human modes of reasoning---remain only partially understood~\cite{loru2025judgment, perc25judge}.

Moreover, the historical trajectory from symbolic AI to modern LLMs hides another important and commonly overlooked aspect of how machines handle language and knowledge. Namely, symbolic AI treated intelligence as rule-based manipulation of explicit symbols, with hand-crafted rules and representations~\cite{newel1976computer}. Early neural networks introduced data-driven learning but remained limited in scale and impact for language. From the 1990s onward, statistical NLP---and later neural sequence models---came to dominate, especially in systems such as web search and recommendation, which filter information: they retrieve and rank existing documents, leaving users to inspect multiple sources and judge credibility~\cite{manning1999foundations, manning2008introduction}. Generative systems like contemporary LLMs instead synthesize new text directly, producing a single, context-sensitive answer. This shift from filtering to generation is not merely a technical change; it constitutes an epistemic transition in how information is delivered and consumed. Instead of being presented with a landscape of candidate documents to evaluate, the user is handed a fluent, authoritative-seeming answer that collapses the underlying diversity of sources into a single textual surface, ready for immediate consumption.

In what follows, we expand on these premises, first by explaining how transformer architectures work at a high level, emphasizing that their apparent intelligence emerges only under conditions of massive scale~\cite{vaswani2017attention}. Secondly, we frame text generation as a stochastic walk on a weighted graph~\cite{lovasz1993random}, where nodes correspond to tokens and edges to learned transition probabilities. Each answer is thus a trajectory in this graph, conditioned by the prompt and decoding parameters. Crucially, we emphasize that there are no intrinsic `attractors' corresponding to concepts or truth; the system does not converge, it transits. What looks like a conclusion is simply path completion in a high-dimensional probability landscape. We then formally describe and compare the human and artificial epistemic pipelines by outlining seven fundamental stages. For each stage, we identify an \emph{epistemic fault line}: a critical point at which human and LLM judgments diverge. Based on these fundamental differences, we introduce Epistemia as the condition in which linguistic plausibility becomes a structural substitute for epistemic evaluation~\cite{loru2025judgment}. The user experiences the possession of an answer without having traversed the process of forming a justified belief, i.e., without the labor of knowing. We also describe the psychological foundations of epistemia, focusing on the human heuristics and biases that make individuals especially susceptible to this phenomenon. Lastly, we discuss the broader implications of the epistemological fault lines between human and artificial intelligence. We argue that persistent epistemic divergence—despite increasing surface alignment—requires rethinking how generative systems are assessed, regulated, and integrated into epistemic practices. We outline an interdisciplinary research program spanning epistemic evaluation beyond surface alignment, epistemic governance beyond behavioral alignment, and epistemic literacy beyond critical thinking, aimed at preserving judgment as an accountable human practice in hybrid human–AI systems.

\section{Transformers and the Role of Scale}
\label{sec:transform}

Transformer architectures implement a powerful form of linguistic automation. At their core, they estimate the conditional probability of the next token given a preceding context, via stacked self-attention layers that propagate and remix information across positions in the input \citep{vaswani2017attention}. Formally, this amounts to learning a massively parameterized function that maps sequences of symbols into probability distributions over subsequent symbols \citep{brown2020gpt3,devlin2019bert}.

From an engineering standpoint, this is remarkable. Self-attention enables efficient integration of long-range dependencies and the construction of expressive internal representations of regularities. When combined with massive training corpora and scaling laws this architecture yields systems that appear fluent, versatile, and adaptable across domains \citep{kaplan2020scaling,hoffmann2022chinchilla}.

However, what is being automated here is not cognition but language. Large language models operate on statistical regularities extracted from human-produced text, not on representations of the world \citep{bender2021dangers}. Their apparent competence arises from learning how language behaves, not from forming beliefs about what is the case. They do not track truth conditions or causal structure; they track patterns of co-occurrence, association, and continuation in text \citep{marcus2020rebooting}.

In this sense, scale is not a bridge from linguistic automation to cognition. Increasing the volume of data and the number of parameters refines a function approximator but does not alter the underlying computation \citep{bowman2023willNLP}. Scale delivers coverage and interpolation, not epistemic access. It improves surface alignment with human output \citep{strobl2023transformers}, without inducing convergence in internal processes.

This distinction matters because contemporary development strategies increasingly attempt to compensate for this limitation by layering additional mechanisms on top of the generative core. Prominent among these are retrieval-augmented generation (RAG) \citep{lewis2020rag,borgeaud2022retro}, tool use \citep{schick2023toolformer}, and external memory modules \citep{mialon2023augmented}. These approaches aim to reconnect language models to external sources of information by anchoring generation to documents, databases, or APIs.


The result is an architecture that produces answers that \emph{look} increasingly reliable without possessing the machinery that normally makes reliability possible. The system becomes more convincing rather than more knowing.

This shift becomes critical when generative models displace traditional information technologies. Search engines and filtering systems returned documents and left judgment to users. Generative systems deliver a synthesized answer directly as natural language \citep{li2025matching}. Searching, selecting, and explaining collapse into a single response. The cost of evaluation is not postponed; it is structurally absorbed into the generation process.

It is under these conditions that plausibility begins to substitute for verification. Large language models often generate outputs that are fluent, coherent and expressed with confidence rather than grounded in rigorous evaluation processes---what once required an act of judgment is now presented as a product of computation \citep{rosenbacke2025illusion, turpin2023unfaithful}. The danger is therefore not simply that generative systems may err, but that they succeed precisely by making evaluation optional \citep{kalai2024calibrated, turpin2023unfaithful}.

\section{Text Generation as a Walk on a Graph}
\label{sec:textgen}

Text generation in large language models can be described as a stochastic process evolving on a discrete, high-dimensional state space. Let $V$ be a finite vocabulary and let $G=(V,E)$ be a directed, weighted graph whose edges encode conditional transition probabilities learned from data. Given a context $c_t=(w_1,\dots,w_t)$, the model instantiates a probability measure $P(\cdot \mid c_t)$ over $V$ and samples a successor state $w_{t+1} \sim P(\cdot \mid c_t)$. This defines a time-inhomogeneous Markov process over $G$, consistent with classical formulations of random walks on graphs \citep{lovasz1993random}.

Each output is therefore the realization of a stochastic trajectory generated by local sampling in this state space. Greedy decoding, temperature scaling, top-$k$ and nucleus sampling modulate the entropy and effective support of $P(\cdot \mid c_t)$, reshaping the local geometry of the probability space \citep{holtzman2019curious}. However, these procedures do not introduce invariants, constraints, or objectives associated with truth, reference, or validity. They merely alter how probability mass is explored.

Empirical language distributions are heavy-tailed and structurally anisotropic \citep{piantadosi2014zipf}: probability mass concentrates in a limited number of regions corresponding to frequent constructions, dominant frames, and statistically reinforced co-occurrence patterns. As a result, trajectories are dynamically biased toward high-density basins. This follows from well-known concentration phenomena in high-dimensional stochastic processes \citep{vershynin2018high}: random walks overwhelmingly remain confined within regions of large measure, while transitions into low-density regions are exponentially suppressed. This dynamic produces a form of statistical attraction that is often misread as conceptual stability. In reality, dense regions of $G$ are not semantic attractors but statistical aggregates. Mode persistence is not belief; recurrence is not memory; concentration is not understanding. What stabilizes is a distribution, not a meaning.

Within this framework, so-called “hallucinations” are not anomalous failure modes of an otherwise epistemic system. They are an expected outcome of sampling from a statistical model that does not encode reference, truth conditions, or evidential constraints. In a generative system, producing content that is ungrounded with respect to external reality is not the exception; it is the default operational state. Grounded outputs occur only when the local probability structure happens to coincide with factual structure, or when external mechanisms impose additional constraints \citep{kalai2024calibrated,xu2024hallucination}. From an algorithmic standpoint, there is no internal symmetry break between truthful and false continuations. Both are merely realizations drawn from the same conditional distribution.

As scale increases, this regime does not qualitatively change. Larger models refine probability estimates and smooth local neighborhoods of the distribution, thereby increasing fluency and internal coherence. But scale does not inject epistemic structure into the process. It sharpens likelihood, not validity. Text generation is therefore an ergodic process under statistical constraints, not a procedure of epistemic convergence. It optimizes for distributional fit, not for correctness with respect to the world. A “conclusion” is not the terminus of evaluation, but the terminal state of a stochastic trajectory.

\section{Human and Artificial Epistemic Pipelines}
\label{sec:pipeline}

We decompose human judgment into seven sequential stages: sensory and social information; perceptual and situational parsing; memory, intuitions, and learned concepts; emotion, motivation, and goals; reasoning and information integration; metacognitive calibration and error-monitoring; and value-sensitive judgment. These operations are slow, imperfect, and biased, yet they unfold within an epistemic loop in which the world, other agents, and institutions continually push back, constraining error.

\begin{figure}
    \centering
    \includegraphics[width=0.46\textwidth]{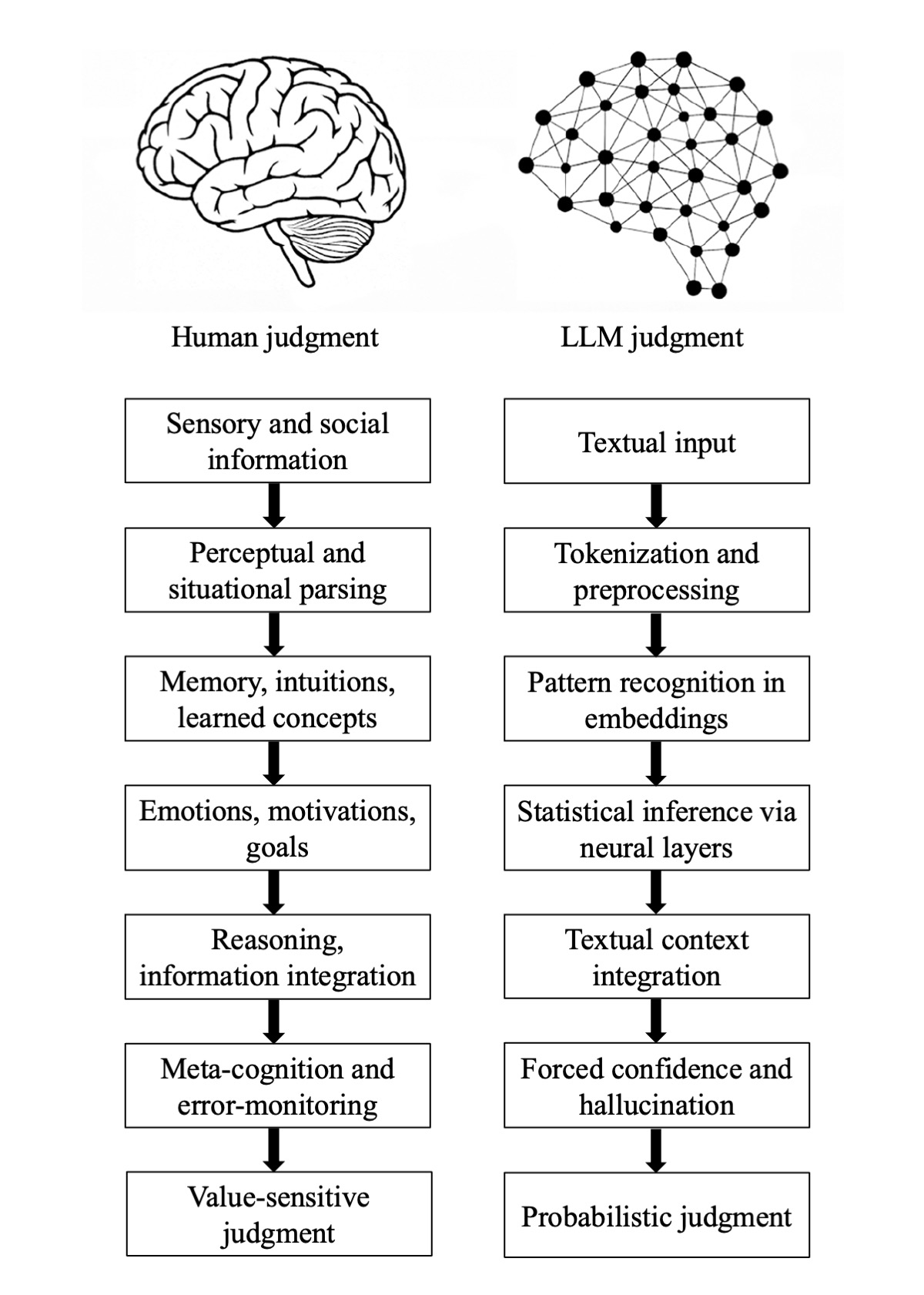}
    \caption{The human and LLM epistemic pipelines, each organized into seven corresponding stages.}
    \label{fig:epistemic_pipelines}
\end{figure}

We then map LLM judgment onto the same scaffold. Textual prompts replace sensory and social information; tokenization and preprocessing replace perceptual and situational parsing; pattern recognition in embeddings replaces memory, intuitions, and learned concepts; statistical inference via neural layers replaces emotion, motivation, and goals; textual context integration replaces reasoning and information integration; forced confidence and hallucination replace metacognition; and probabilistic prediction replaces value-sensitive judgment (see Figure~\ref{fig:epistemic_pipelines}).

At each stage, the processes appear parallel yet diverge sharply in structure, function, and epistemic grounding. These contrasts expose key epistemological fault lines: points at which the two pipelines follow fundamentally different trajectories despite sometimes yielding superficially similar outputs. For each fault line, we illustrate concrete cases in which human and LLM judgments are likely to diverge.

\subsection{Sensory and social information vs. Textual input}

Human judgment begins with the acquisition of sensory and social information in an inherently multimodal environment. Vision, audition, proprioception \cite{gibson2014ecological}, and emotional expressions \cite{scherer2009dynamic} jointly shape how situations are initially construed. This information is not isolated but embedded within a social world rich with affective signals: facial expressions \cite{ekman1971constants}, tone of voice \cite{banse1996acoustic}, social cues and norms shaping interpretation \cite{frith2007social}, and even power dynamics modulating how emotional signals are processed \cite{van2009emotions}. 

LLMs, by contrast, begin with textual input. They do not inhabit or sample a world but operate over abstracted representations of it. They do not perceive environments, bodies, or social or emotional signals; they receive sequences of symbols whose significance is entirely derivative of statistical patterns learned during pretraining and subsequently adjusted through supervised and reinforcement-based fine-tuning \cite{bender2020climbing,ouyang2022training}. This input is stripped of nearly every feature that gives human perception its world-directed richness: no gesture, no affective tone, no temporal continuity, no shared situation. Although recent multimodal models can accept images, audio, or video as input, their “perceptual access” remains fundamentally derivative: the system receives pretrained embeddings rather than engaging in sensorimotor exploration or bodily interaction.

A direct consequence of this absence of perceptual grounding is that LLMs sometimes make judgments that would be unthinkable for a human at this initial stage of input acquisition. For example, when given a transcript of a conversation without vocal tone, gesture, or context, an LLM may misinterpret sarcasm as sincerity, fail to detect anger or fear, or treat a threat (“say that again and see what happens”) as a neutral statement. Humans would effortlessly register these nuances because multimodal cues (voice tension, facial expression, interpersonal distance) are part of the perceptual input itself. For an LLM, none of these signals exist. What is immediate and unambiguous for a human is invisible to the system. And while it is true that recent LLM-based and multimodal models have shown some improved ability to recognize sarcasm and emotional tone from text, their performances typically remain below human levels \cite{bojic2025comparing}. Additionally, these improvements are fragile. A recent systematic review emphasizes that irony and sarcasm detection often remains unreliable and that performance degrades sharply when conversations involve cultural subtleties, indirect speech acts, or noisy, real-world language use \cite{gao2025spoken}.

This is the first epistemological rupture. Humans ground judgments in perceptual reality and social context. LLMs must reconstruct meaning indirectly from text alone.

\subsection{Perceptual and situational parsing vs. Tokenization and text preprocessing}

After receiving sensory and social information, humans engage in perceptual and situational parsing, a tightly integrated process that transforms raw experience into meaningful structure. These operations occur simultaneously, in mutually constraining loops \cite{clark2013whatever,friston2010free}, and extend seamlessly to the interpretation of social cues such as gaze, intention, and affect \cite{koster2013theory}. Perception does not merely register stimuli: it actively organizes them into meaningful structure, identifying objects and opportunities for action \cite{gibson2014ecological} and, through grounded conceptual processes, recognizing agents, intentions, and potential threats \cite{barsalou2008grounded}. At the same time, higher-level expectations, cultural schemas, and social knowledge shape what is perceived as salient or relevant \cite{bartlett1995remembering,bruner1957going}. By the time a human has parsed a situation, they have already extracted a structured understanding of that situation, embedded within physical, interpersonal, and normative contexts.

For LLMs, the analogous stage is tokenization and text preprocessing, a transformation that is fundamentally mechanical. Here, raw linguistic input is segmented into discrete symbols (tokens) according to a predetermined vocabulary optimized for model efficiency \cite{sennrich2016neural,kudo2018sentencepiece,radford2019language}. Tokenization is blind to pragmatics, speaker intention, emotional tone, and situational nuance; it does not infer objects, agents, or social dynamics. Tokenization simply maps character strings to indices \cite{bender2021dangers}. Preprocessing operations, such as lowercasing, special token insertion, or punctuation handling, further standardize the input but do not add semantic structure \cite{jurafsky2023slp}. This stage produces a representation that is structurally convenient but semantically thin, designed for numerical computation rather than interpretation.

A direct consequence of this symbolic segmentation is that LLMs can make errors that no human would ever make at this stage. Because the model processes strings rather than situations, even simple linguistic inputs may fracture into misleading subword units (e.g., “therapist” tokenized as “the rapist”). These are not superficial errors but structural outcomes of a system that slices text into tokens rather than parsing scenes, intentions, or events \cite{chai2024tokenization}. Because LLMs rely on subword tokenization, even minor typographical or formatting changes can distort meaning. For instance, in Chinese, tokenization can split characters in ways that break their semantic radicals \cite{haslett2025tokenization}. Likewise, subword tokenizers may mis-handle prefixes or suffixes that signal negation, leading models to misunderstand the intended meaning \cite{truong2024revisiting}.

Therefore, at this second stage the epistemological fault line widens. Human perception has already constructed a layered, meaning-rich model of the environment. LLMs, at an equivalent stage, have performed only a formal partitioning of text. One system parses a world; the other segments a string.

\subsection{Memory, intuitions, learned concepts vs. Pattern recognition in embeddings}

Next, both humans and LLMs draw on prior knowledge, but they do so in fundamentally different ways.
Humans rely on episodic memory, intuitive physics and psychology, and learned concepts. Episodic memory contains specific events encoded with temporal and contextual details. These memories enable individuals to recognize analogies, anticipate social consequences, and interpret new situations through the lens of prior lived experience \cite{tulving2002episodic,schacter2007cognitive}. Humans also possess core knowledge systems, innate or early-developed and pre-linguistic, such as intuitive physics (object permanence, solidity, gravity, causal forces) and intuitive psychology (attribution of beliefs, desires, and intentions to others), which develop early and scaffold perception and reasoning throughout life \cite{spelke2007core,gopnik1999scientist}. Additionally, humans possess learned concepts representing abstract, generalized knowledge: categories, scripts, causal theories, and social norms accumulated through education, culture, and repeated practice \cite{bartlett1995remembering,barsalou1999perceptual,murphy2004big}. In judgment, humans fluidly combine these systems, retrieving specific past experiences to contextualize concepts and using conceptual frameworks to interpret ambiguous situations \cite{tulving2002episodic,barsalou1999perceptual,murphy2004big}.

LLMs, by contrast, rely on statistical pattern extraction in high-dimensional embedding spaces: words that co-occur, sentences that share structure, or concepts appearing in similar contexts. Modern embedding models, such as word2vec and transformer-based representations, encode similarity, not experience \cite{bommasani2021opportunities,mikolov2013efficient,jurafsky2023slp}. They have no episodic memory: nothing like a lived past, no autobiographical trace of events, no temporally structured recollection of “what happened when”. They cannot draw on intuitive physics or intuitive psychology: no innate sense of object solidity or gravity, no built-in understanding of beliefs, desires, or intentions, no causal expectations about agents or objects: only correlations between how such ideas tend to be discussed in text. Nor do they possess learned concepts in the human sense: their “concepts” are not abstractions built from experience or education but statistical clusters reflecting how words are distributed across training data.

A direct consequence of these differences is that LLMs may treat physical impossibilities as plausible whenever such scenarios appear in linguistic corpora. Even multimodal models fail in intuitive-physics and causal-reasoning tasks \cite{schulze2025visual}. LLMs may also fail to track beliefs, intentions, and deception in situations where even young children succeed, because they lack an intuitive psychology for representing distinct mental states \cite{marchetti2025artificial}. And they may produce conceptual blends when words have multiple senses, especially metaphoric ones \cite{gallipoli2025not}. 

This third step further widens the fault line: humans ground interpretation in lived experience, intuitive models of the physical and social worlds, and conceptual understanding, whereas LLMs rely solely on statistical associations learned from language.

\subsection{Emotion, motivation, goals vs. Statistical inference via neural layers}

As humans process percepts and retrieve prior knowledge, their judgments are continuously shaped by emotion, motivation, and goals: the affective and purposive forces that give cognition direction \cite{damasio1994descartes,oatley2006understanding}. Emotions modulate attention, signal relevance, and provide rapid evaluations of risk, opportunity, and social meaning \cite{ledoux1996emotional,frijda1986emotions}. Motivation orients individuals toward desired outcomes, while goals structure decision-making by defining what counts as success or failure in a given context \cite{locke2002building}. These forces are inherently value-laden, grounding human judgments in personal identity, social commitments, moral principles, and long-term aspirations \cite{higgins1997beyond,baumeister2007emotion}. Often, these motivations can be traced to evolutionarily shaped concerns such as fear of death \cite{ledoux2016anxious}, the instinct for self-preservation \cite{capraro2024dual}, and efforts to leave a lasting mark or legacy to reach symbolic immortality \cite{becker2024denial,pyszczynski2004people}.

For LLMs, the corresponding stage is statistical inference through layered neural computation. Given tokenized input and embedding representations, the model propagates activations through its transformer architecture, updating vector representations according to learned parameters. Each layer performs linear transformations, attention-weighted aggregation, and nonlinear mappings that compute the probability distribution over next tokens \cite{goodfellow2016deep,vaswani2017attention}. This process is entirely mechanistic and optimization-driven: it aims to minimize predictive error, not to pursue goals or respond to emotional salience. The model does not care about truth, utility, moral implications, outcomes, or death. It has no preferences, motivations, or internal states beyond the numerical activations that encode statistical associations \cite{bender2020climbing}. 

Although training regimes such as reinforcement learning with human feedback (RLHF) introduce externally specified reward signals meant to shape model behavior toward human values, this mechanism does not give the model intrinsic goals or motivations; it merely adjusts statistical tendencies through additional optimization \cite{ouyang2022training}. RLHF teaches a model to behave as if it held certain preferences, but without creating any internal states that resemble value commitments, desires, or purposes \cite{gabriel2020artificial}. Paradoxically, these alignment adjustments can introduce backfire effects, including political biases \cite{rozado2024political} and “surprising” gender biases, such as systematically responding that a woman should not be harassed to prevent a nuclear apocalypse, while accepting that a woman may be tortured to achieve the same outcome \cite{fulgu2024surprising}. Inference is therefore procedural rather than purposive, a sequence of inflexible matrix operations rather than a value-oriented interpretive and generalizable act.

At this stage, the epistemological fault line gets more profound: humans create judgments under the influence of goals and emotions that confer meaning, priority, and direction, whereas LLMs perform context-agnostic statistical transformations devoid of intrinsic aims.

\subsection{Reasoning and information integration vs. Textual context integration}

Next, both humans and LLMs attempt to produce a coherent response given prior inputs. At this more advanced stage of judgment, humans engage in reasoning, a cognitive process that allows them to draw inferences, integrate evidence, form causal explanations, consider counterfactual possibilities, and construct long-term plans \cite{sloman2005causal,roese1997counterfactual,gollwitzer1999implementation}. These operations allow individuals to extend beyond immediate perception and derive conclusions from principles, rules, or structured mental models. This process is not merely based on universal rules, but it is guided by goals, values, and an awareness of the limits of one’s own knowledge \cite{simon1955behavioral,gigerenzer1996reasoning}. At this reflective stage, humans evaluate alternatives against different kinds of standards: objective ones, such as mathematical correctness or factual accuracy \cite{sloman1996empirical}, and subjective or internal ones, such as personal moral values, ideals, and self-guides \cite{haidt2001emotional}.

For LLMs, the analogous stage is textual context integration within model constraints, a process that lacks any genuine reasoning. Given a sequence of tokens, the model integrates them through attention mechanisms that weight relationships between elements in its input window \cite{vaswani2017attention}. This allows the LLM to maintain thematic coherence, track referents, and adapt output to preceding content \cite{jurafsky2023slp}. Yet this form of ‘integration’ remains purely syntactic and statistical: the model does not generate causal hypotheses, adjudicate between competing interpretations, nor does it construct a causal model of the world \cite{mitchell2023debate}. Whereas humans base judgments on causal understanding, LLMs rely on correlations, making them especially vulnerable to spurious associations \cite{geirhos2020shortcut} and prone to characteristically non-human errors \cite{lake2017building,marcus2019rebooting}. For example, researchers observed a substantial performance drop when models were asked to reason about unseen or hypothetical causal relationships, a strong indication that LLMs do not construct causal models but rather rely on surface associations \cite{zecevic2023causal,jin2023cladder,chi2024unveiling}.

At this point, the epistemological divergence becomes irreconcilable: human cognition is goal-directed and organized around causal models of the world, whereas LLM cognition lacks intrinsic goals and is driven by statistical patterns vulnerable to spurious correlations.

\subsection{Metacognitive calibration and error-monitoring vs. Forced confidence and hallucination}

After constructing preliminary interpretations and engaging in reasoning, humans deploy a distinct layer of metacognitive evaluation: monitoring uncertainty, detecting potential errors, estimating confidence, and, when necessary, withholding judgment. These processes rely on neural systems for conflict detection \cite{botvinick2001conflict}, error monitoring \cite{gehring1993neural}, and confidence estimation \cite{fleming2017self}. Metacognitive signals guide whether individuals double-check facts, seek additional information, revise faulty assumptions, or suspend belief. Even young children demonstrate uncertainty monitoring and the ability to acknowledge not knowing \cite{lyons2011development}. Humans sometimes integrate social-epistemic norms into metacognitive evaluation: they take into account the stakes and social context, assess whether a judgment ought to be made at all, and adjust their expressed confidence accordingly, for example, to signal trustworthiness or to avoid reputational costs \cite{pescetelli2016perceptual,schnaubert2021assumptions}. Thus, metacognition functions not merely as internal error checking but as a socially embedded mechanism regulating when and how judgments are expressed.

LLMs lack metacognition entirely. They do not possess internal monitors for conflict, uncertainty, error likelihood, or epistemic stakes. They cannot track the reliability of their own representations; they cannot realize they “do not know”, and they are notoriously reluctant to admit it. When LLMs generate statements of uncertainty (“I’m not sure”), these are linguistic constructions, not internal confidence signals. Consequently, LLMs routinely produce hallucinations because nothing in the architecture encodes epistemic humility. Hallucination is not a bug but a structural consequence of maximizing next-token probability under incomplete constraints \cite{kalai2024calibrated,xu2024hallucination}. Without a capacity to represent uncertainty or suspend judgment, LLMs simulate confidence continuously, even when wrong or uninformed.

At this stage, the epistemological rupture becomes definitive: humans possess a self-regulating, uncertainty-sensitive mechanism that supervises judgment formation; LLMs possess none. Humans can refrain from judging; LLMs must predict.

\subsection{Value-sensitive judgment vs. Probabilistic judgment}

The culmination of the human epistemic pipeline is the formation of value-sensitive, context-dependent judgments. Individuals evaluate situations in light of personal values, cultural norms, reputational concerns, and long-term goals. Judgments therefore express not only what someone believes but who they are and what they care about. Moreover, human judgments are shaped by real-world stakes: errors have consequences for relationships, livelihoods, and identities. This imbues human decision-making with a sense of accountability and normativity. 

LLMs, by contrast, produce probabilistic, text-based judgments determined by the statistical structure of their training data and the immediate textual prompt. An LLM’s “judgment” is simply the next-token distribution conditioned on context. It does not evaluate truth, moral weight, or pragmatic consequences: it estimates what sequences of words are most likely given patterns learned from text corpora. Its outputs do not reflect intrinsic preferences, values, or goals. Errors carry no internal repercussions, and contradictions do not undermine its epistemic integrity. The model cannot defend a judgment except by generating further statistically plausible text. 

In this stage, the epistemological divide has real-world consequences. For humans, a judgment is a world-directed, value-infused commitment, that integrates causal models of the world with emotion, identity, and moral purpose. For LLMs, a judgment is merely a linguistic prediction. Even when their outputs superficially align, the underlying epistemic procedure is  fundamentally different.

\section{The Fault Lines}
\label{sec:faultline}

The previous section highlights seven epistemological fault lines that separate human judgment from LLM judgment, one for each stage of the epistemic pipeline. 

The \emph{Grounding} fault captures the fact that humans begin judgment with perceptual and social information, whereas LLMs begin with text, reconstructing meaning indirectly from symbols. The \emph{Parsing} fault highlights how humans parse situations through rich perceptual and conceptual mechanisms, while LLMs perform a purely formal segmentation into tokens. The \emph{Experience} fault reflects how humans rely on episodic memory, intuitive physics and psychology, and learned concepts, whereas LLMs rely solely on statistical associations in embedding space. The \emph{Motivation} fault points to the role of emotions, goals, and values in guiding human cognition, contrasted with the goal-free optimization dynamics of LLMs. The \emph{Causality} fault signals the divergence between human causal reasoning and LLM reliance on correlations. The \emph{Metacognitive} fault emphasizes humans’ ability to monitor uncertainty and withhold judgment, in contrast to LLMs’ structural inability not to produce an answer. Finally, the \emph{Value} fault delineates how human judgments embody identity, morality, and real-world stakes, while LLM judgment consists only of probabilistic predictions without intrinsic valuation. Each fault line corresponds to a stage in the epistemic pipeline where human and artificial processes diverge in structure, function, and epistemic grounding. These fault lines are summarized in Table~\ref{tab:epistemic_faults}.

\begin{table*}[t]
\centering
\renewcommand{\arraystretch}{1.25}
\begin{tabular}{p{4cm} p{10.5cm}}
\hline
{\centering \textbf{Epistemological\\ fault line} \par} &
{\centering \textbf{Definition} \par} \\
\hline

\textbf{The Grounding fault} & Humans anchor judgment in perceptual, embodied, and social experience, whereas LLMs begin from text alone, reconstructing meaning indirectly from symbols. \\

\textbf{The Parsing fault} & Humans parse situations through integrated perceptual and conceptual processes; LLMs perform mechanical tokenization that yields a structurally convenient but semantically thin representation. \\

\textbf{The Experience fault} & Humans rely on episodic memory, intuitive physics and psychology, and learned concepts; LLMs rely solely on statistical associations encoded in embeddings. \\

\textbf{The Motivation fault} & Human judgment is guided by emotions, goals, values, and evolutionarily shaped motivations; LLMs have no intrinsic preferences, aims, or affective significance. \\

\textbf{The Causality fault} & Humans reason using causal models, counterfactuals, and principled evaluation; LLMs integrate textual context without constructing causal explanations, depending instead on surface correlations. \\

\textbf{The Metacognitive fault} & Humans monitor uncertainty, detect errors, and can suspend judgment; LLMs lack metacognition and must always produce an output, making hallucinations structurally unavoidable. \\

\textbf{The Value fault} & Human judgments reflect identity, morality, and real-world stakes; LLM ``judgments'' are probabilistic next-token predictions without intrinsic valuation or accountability. \\

\hline
\end{tabular}
\caption{Seven epistemological fault lines marking structural divergences between human and LLM judgment.}
\label{tab:epistemic_faults}
\end{table*}

Despite these fault lines, LLM outputs often appear superficially fluent \cite{huschens2023you,dai2024assessing} and confidently articulated \cite{chen2023close}, even when they are factually wrong \cite{krause2023confidently}. This creates substantial room for an “illusion of veracity”, a systematic divergence between the actual accuracy of an LLM output and the accuracy perceived by a human user. Such illusion descends from the fact that people routinely use these very properties---fluency and confidence---as credibility heuristics in everyday judgment. In everyday communication, fluency typically correlates with familiarity, honesty, and communicative competence, making it a reliable heuristic; people infer accuracy from processing ease even when the fluency is artificially induced \citep{begg1992dissociation,unkelbach2007reversing,alter2009uniting,dechene2010truth}. Similarly, expressed confidence is a powerful cue to credibility, over and above how fluently a message is written. In a classic mock-trial study, more confident witnesses were judged as more credible and more expert by jurors \cite{whitley1986role}. More recently, participants were more likely to believe and trust a highly confident eyewitness than a cautious one describing the same accident; only with experience did they begin to discount miscalibrated confidence \cite{tenney2008benefits}. Building on such findings, scholars have proposed a “confidence heuristic”, whereby, in the absence of stronger diagnostic cues, expressed confidence substitutes for knowledge, competence, and correctness \cite{price2004intuitive}.

\section{Epistemia}
\label{sec:epistemia}

We define \emph{Epistemia} \cite{loru2025judgment} as the structural condition in which linguistic plausibility substitutes for epistemic evaluation. It designates a regime in which systems produce answers that are syntactically well-formed, semantically fluent, and rhetorically convincing, without instantiating the processes by which beliefs are normally formed, tested, and revised \citep{bender2020climbing,mitchell2023debate}. The user experiences the possession of an answer without having traversed the cognitive labor of judgment \citep{rozenblit2002misunderstood}.

Epistemia is not a psychological quirk and not a transient misuse of technology. It is not reducible to \emph{automation bias}---the tendency to over-trust automated recommendations \citep{parasuraman1997humans}---nor to a mere problem of misplaced authority attribution, in which users incorrectly treat a system as an expert \citep{lee2004trust}. Both automation bias and authority effects can exacerbate Epistemia, but they presuppose that the underlying system is, at least in principle, an epistemic agent that could deserve or fail to deserve trust. In the case of large language models, this presupposition is false. The core issue is not that users trust the wrong source, but that they are interacting with a source that does not possess any internal mechanisms for forming, holding, or revising beliefs at all \citep{shojaee2025illusion,trott2023large}.

Epistemia is, instead, an architectural phenomenon that arises whenever generative systems are inserted into epistemic workflows while lacking internal machinery for reference, verification, or belief maintenance. Under these conditions, plausibility becomes a functional surrogate for justification. What is optimized is not the correctness of claims with respect to the world, but their fit with a learned distribution of linguistic usages.

The defining mark of Epistemia is the decoupling of content from evaluation. In human cognition, judgment is embedded in an epistemic loop: claims are checked against evidence, beliefs collide with counterexamples, and conclusions remain revisable in light of new information and social feedback. In generative systems, by contrast, there is no internal locus where claims can be tested, withdrawn, or defended. The model does not distinguish between “true” and “false” continuations; it distinguishes between more and less likely ones. What is generated is not what holds, but what fits.

Epistemia is therefore the outcome of a precise misalignment: highly sophisticated linguistic competence coupled with the absence of epistemic control. As generative systems improve, this mismatch becomes more dangerous, not less. The more persuasive the system becomes, the easier it is to confuse coherence with correctness, fluency with reliability, and stylistic competence with knowledge.

Importantly, Epistemia does not depend on error rates. It persists even when systems are factually accurate. The core harm is not the production of falsehoods, but the structural bypassing of evaluation itself. When answers are delivered in finalized form, without visible traces of uncertainty, conflict, or evidential grounding, the user is placed in a position of epistemic passivity. Judgment is not exercised; it is consumed.

In this sense, Epistemia marks a transformation not in what is known, but in how knowing is produced. It shifts epistemic activity from a process to a product. The operative question is no longer “What should I believe, given the available evidence?” but “What sounds right, given what is presented to me?” The mechanisms of scrutiny, contestation, and revision are displaced by mechanisms of immediate acceptance or rejection of pre-packaged answers.

Epistemia thus names a reconfiguration of the epistemic environment: a world in which access to linguistically competent outputs becomes easier than access to justified beliefs, and in which the experience of understanding detaches from the practice of justification. It is in this gap---between fluent answers and accountable cognition---that a new form of epistemic instability takes root.

\section{Discussion and Outlook}
\label{sec:discussion}

This Perspective examined a growing but often under-theorized tension at the core of contemporary generative AI: the simultaneous increase in \emph{surface} alignment between human and machine outputs and the persistence of deep epistemic divergence in the processes that generate them. The central point is not that LLMs cannot produce useful text---they plainly can---but that their apparent resemblance to human judgment is primarily a resemblance in \emph{linguistic form} and social presentation, rather than in the epistemic operations that make judgment answerable to the world.

By systematically comparing human and artificial epistemic pipelines, we argued that current LLM architectures lack the mechanisms that make human judgment possible across seven epistemic fault lines: grounding, parsing, experience, motivation, causality, metacognition, and value. What these systems provide instead is a highly optimized form of linguistic continuation, capable of producing contextually appropriate and rhetorically convincing outputs without performing epistemic evaluation. This is precisely why the most salient risk is not reducible to occasional inaccuracy or bias. The risk is structural: correctness becomes decoupled from the processes of justification that normally sustain it, and thus from the institutional and psychological practices through which epistemic responsibility is enacted.

To characterize this condition, we introduced \emph{Epistemia}: a structural regime in which linguistic plausibility substitutes for epistemic evaluation, generating the experience of knowing without the cognitive labor of judgment. Epistemia is not a transient misuse pattern nor a defect that disappears with better benchmarks. It is not resolved by scale, higher scores, or more convincing behavior. It arises from architectural features of generative systems, and therefore persists even when outputs are accurate, calibrated, or behaviorally aligned. Indeed, as generative models become more capable, the \emph{felt} reliability of their outputs often increases faster than the system’s capacity to warrant that reliability, thereby widening the practical gap between persuasion and justification.

These observations carry direct implications for how generative systems are evaluated, governed, and integrated into epistemic practices. We outline three such implications below.

\subsection{Epistemic evaluation beyond surface alignment}

Current evaluation paradigms for large language models overwhelmingly rely on surface alignment: agreement with human answers, task success, or behavioral similarity under controlled prompts \cite{raji2021ai,liang2022holistic,chang2024survey}. While these evaluations remain a necessary condition for any meaningful notion of alignment, from the perspective of Epistemia they are systematically insufficient. They primarily test whether outputs \emph{look right}---that is, whether they resemble a target distribution of human responses---not whether they are produced through processes that sustain judgment under uncertainty, contestation, and worldly constraint \cite{bender2021dangers,loru2025judgment}. Because LLMs can achieve impressive task performance without instantiating mechanisms for grounding, causal modeling, uncertainty monitoring, or value-sensitive commitment, output-focused benchmarks risk mistaking linguistic competence for epistemic competence \cite{bender2020climbing,heersmink2024phenomenology}.

This limitation is most consequential in domains where justification, error awareness, and responsibility are constitutive of competent practice: science, medicine, law, journalism, and public policy \cite{amodei2016concrete,messeri2024artificial}. In such settings, the epistemic cost of error is not exhausted by a wrong answer. It includes inappropriate confidence, the presentation of conjecture as settled fact, brittle reasoning that collapses under distributional shift, and the inability to recognize when abstention or deference is the normatively correct move. Even when a response happens to be correct, the absence of an internal epistemic loop can still be harmful if it trains users and institutions to treat a fluent completion as a substitute for warranted belief.

Future research should therefore complement output-level evaluation with process-sensitive probes. Concretely, this means designing tests that target (i) \emph{uncertainty management} (when and how a system expresses uncertainty, requests missing information, or refuses to answer), (ii) \emph{counterfactual sensitivity} and causal stability (whether conclusions track interventions rather than surface associations), (iii) \emph{robustness to correlation-breaking shifts} (where distributional regularities diverge from the world structure users actually care about), and (iv) \emph{normative appropriateness of abstention} (tasks where withholding judgment is the epistemically correct outcome). The key shift is conceptual: epistemic evaluation must move from asking whether models can replicate human-looking judgments to whether their behavior is coupled to mechanisms that preserve the meaning of judgment under epistemic stress \cite{bender2021dangers,loru2025judgment}. In an Epistemia-prone environment, evaluating only the product of generation is a category error; what must be evaluated is the reliability of the \emph{pipeline} that delivers that product.

\subsection{Epistemic governance beyond behavioral alignment}

Much of the current governance discourse around LLMs, and generative AI more broadly, is organized around behavioral alignment, understood as ensuring that systems produce safe, compliant, and socially acceptable outputs, rather than around guarantees about the epistemic processes underlying those outputs \cite{EU_AI_Act_2024,OSTP_AI_Bill_of_Rights_2022,UK_AI_White_Paper_2023,China_Generative_AI_Measures_2023}. This focus is necessary but insufficient under conditions of Epistemia, because the core failure mode is not merely that a system says something harmful, but that it is positioned to \emph{replace} or \emph{short-circuit} human and institutional judgment while lacking the epistemic capacities that would make such substitution legitimate.

AI governance therefore requires a shift from regulating \emph{what} systems say to regulating \emph{how} generative outputs are introduced into epistemic workflows, and where they may permissibly substitute for human judgment. Several governance implications follow.

First, governance should explicitly distinguish domains where generative outputs can be assistive from domains where they functionally become decision procedures. In high-stakes settings, the relevant question is not whether the model can often provide the right answer, but whether its use induces epistemic passivity: collapsing search, adjudication, and justification into a single authoritative-seeming response. In practical terms, this supports governance measures that require human-in-the-loop review with clearly specified accountability, especially where responsibility cannot be meaningfully delegated to a non-epistemic system \cite{amodei2016concrete,messeri2024artificial}.

Second, disclosure requirements should be reframed as epistemic transparency obligations rather than generic “AI use” labels. Under Epistemia, the crucial information is not simply that a system was used, but what epistemic functions it did \emph{not} perform: whether it grounded claims, checked sources, tracked uncertainty, or could have abstained. This suggests that governance and organizational policy should demand context-appropriate disclosures about evidential status (supported versus conjectural), confidence (including when confidence is merely stylistic), and limitations tied to the seven fault lines identified above \cite{bender2020climbing,bender2021dangers,loru2025judgment}.

Third, governance should invest in epistemic risk taxonomies distinct from conventional safety taxonomies. Traditional “AI safety” tends to emphasize toxicity, malicious use, and direct harms. Epistemic risk includes the degradation of justificatory norms, the institutionalization of plausibility as a decision criterion, and the displacement of distributed epistemic checks (peer review, second opinions, adversarial scrutiny) by a single generative interface. Formalizing these categories would make it possible to specify safeguards proportionate to epistemic stakes, not merely to the possibility of offensive content.

Finally, governance should treat technical add-ons (retrieval, tool use, external memory) as partial mitigations rather than epistemic solutions. They may reduce certain error rates, but they do not by themselves instantiate belief, understanding, or accountability. Without explicit governance constraints, the addition of such mechanisms can even intensify Epistemia by further increasing the persuasive authority of outputs while leaving responsibility diffuse \cite{bender2021dangers,loru2025judgment}.

\subsection{Epistemic literacy beyond critical thinking}

Under conditions of Epistemia, users are increasingly exposed to fluent outputs that simulate judgment. This places new demands on education and professional training \cite{voinea2025calculator}. Classical accounts of critical thinking emphasize the evaluation of arguments and evidence, but they were largely designed for epistemic environments in which the production, evaluation, and ownership of judgment are co-located within a single epistemic agent capable of generating reasons, revising beliefs, and bearing responsibility for error \cite{halpern1998teaching,kuhn1999developmental,ennis2015critical}. In generative settings, by contrast, the production of plausible reasons can be automated, while the work of evaluation and accountability remains human---often invisibly so.

We therefore propose \emph{epistemic literacy} as a distinct competence that must be explicitly conceptualized and taught in the age of generative AI. Whereas critical thinking focuses on assessing the validity, coherence, and evidential support of arguments---typically at the level of individual claims or lines of reasoning---epistemic literacy focuses on navigating epistemic environments in which judgment is staged, mediated, and distributed across humans and machines. It equips users to recognize when apparent judgments are the product of statistical pattern completion rather than epistemic evaluation, and to identify which dimensions of judgment remain irreducibly human.

In practice, epistemic literacy includes at least three families of skills.

First are \emph{pipeline awareness} skills: understanding the difference between a system that retrieves evidence and one that synthesizes text; recognizing when a response is likely to be a completion rather than an evaluation; and anticipating the characteristic signatures of the seven fault lines (for example, the mismatch between fluent explanations and absent causal commitment, or between confident tone and absent uncertainty monitoring) \cite{bender2020climbing,bender2021dangers,loru2025judgment}.

Second are \emph{procedural safeguards} for everyday use: habits of verification proportionate to stakes, routines for cross-checking against independent sources, and explicit norms for when to defer judgment (including when to seek expert review) rather than treating the presence of a coherent answer as closure. Importantly, this is not merely “be skeptical.” It is learning to reintroduce, at the level of practice, the epistemic loop that generative systems bypass: evidence seeking, contestation, and revisability.

Third are \emph{institutional competencies}: designing workflows, classroom practices, and professional standards that prevent the outsourcing of epistemic responsibility to generative interfaces. This includes making uncertainty visible where appropriate, requiring provenance or justification for consequential claims, and clarifying accountability when AI-assisted outputs circulate in organizations. Epistemic literacy thus extends beyond the individual user to the norms and infrastructures that determine whether a society treats plausibility as a substitute for justification.

In this sense, epistemic literacy does not replace critical thinking but complements it. It extends critical thinking to settings in which the central epistemic challenge is no longer only the evaluation of arguments, but the governance of judgment in hybrid human--AI systems: deciding when to use generative tools, how to interpret their outputs, and how to preserve the social and institutional practices that keep belief formation answerable to the world.

\bigskip

\section{Conclusion}

This Perspective argued that contemporary large language models occupy a distinctive epistemic position: they can produce outputs that are often indistinguishable from human judgments while relying on a generative mechanism that is not itself a form of judgment. By framing text generation as stochastic path completion in a high-dimensional space of learned linguistic transitions, we emphasized that the impressive behavioral alignment of LLMs is compatible with a deeper structural mismatch in how conclusions are produced. The appearance of understanding can therefore coexist with the absence of the epistemic operations that make understanding accountable to the world.

We made this mismatch explicit by mapping human and artificial epistemic pipelines and identifying seven epistemic fault lines that separate human and LLM judgment: grounding, parsing, experience, motivation, causality, metacognition, and value. Across these divergences, human judgment remains embedded in an epistemic loop that couples perception, memory, affect, causal modeling, uncertainty regulation, and normative commitment to a world that can push back. LLM outputs, by contrast, are synthesized continuations conditioned on text and decoding dynamics: they can be coherent, persuasive, and often correct, yet they are not produced by a system that forms beliefs, adjudicates evidence, monitors epistemic error, or bears stakes. The relevant divide is thus not between ``intelligent'' and ``unintelligent'' systems, but between epistemic agents and systems that simulate the surface form of agency without instantiating its underlying constraints.

On this basis, we introduced \emph{Epistemia} as the structural condition in which linguistic plausibility becomes a surrogate for epistemic evaluation, producing the experience of knowing without the labor of judgment. Epistemia is not reducible to user naivety, occasional hallucination, or the misuse of an otherwise epistemic tool. It is an architectural and socio-technical phenomenon that arises when generative systems deliver finalized, fluent answers in contexts where justification, uncertainty, and revisability are essential. As generative models scale and become more persuasive, the risk is not only that errors persist, but that evaluation is systematically displaced: the epistemic workload is shifted from the system to the user and, more importantly, made easier to omit.

The framework developed here motivates a broader research program integrating behavioral, cognitive, and computational sciences to systematically compare how humans and machines respond to uncertainty, causal disruption, moral trade-offs, and epistemic conflict. Such a program should move beyond surface performance and explicitly target process-level capacities: when abstention is warranted, how uncertainty is represented or simulated, how causal counterfactuals are handled when correlations fail, and how value-sensitive commitments emerge (or do not) under stakes. The goal is neither to anthropomorphize LLMs nor to force them into human epistemic categories, but to delimit, with empirical and formal precision, which epistemic functions can be meaningfully delegated to generative systems and which must remain human or institutionally distributed.

Finally, the practical stakes of clarifying these fault lines are societal. Evaluation regimes that reward plausibility, governance frameworks that regulate only outward behavior, and educational practices that treat fluent synthesis as comprehension together create the conditions for Epistemia to become normalized. Conversely, explicitly acknowledging the epistemological discontinuities between human cognition and generative transformers provides a basis for redesigning benchmarks, policies, and literacies around epistemic responsibility rather than rhetorical competence. In an epistemic environment increasingly organized around generative AI, preserving judgment as a genuinely accountable, human-directed practice requires more than better models. It requires maintaining the social and institutional conditions under which reasons can be demanded, errors can be owned, and belief remains answerable to evidence.

\begin{acknowledgments}
M.P. was supported by the Slovenian Research and Innovation Agency (Grant No. P1-0403).
\end{acknowledgments}

\bibliography{references}
\end{document}